\documentclass{aa}  

\usepackage{graphicx}
\usepackage{txfonts}
\usepackage{natbib}
\usepackage{xcolor}
\bibpunct{(}{)}{;}{a}{}{,}
\usepackage{xcolor}
\usepackage{rotating}
\usepackage{array}
\usepackage{pdflscape}

\usepackage{subfig}

\usepackage{threeparttable}

\begin{document}

\title{Investigation into the origin of the soft excess in Ark 564 using principal component analysis}

\author{Ming\ Lyu  \inst{1,2}
\and Zhenyan \ Fei \inst{1,2} 
\and Guobao \ Zhang \inst{3,4}
\and X. J. Yang \inst{1,2}
 }

\institute{Department of Physics, Xiangtan University, Xiangtan, Hunan 411105, China
\and
Key Laboratory of Stars and Interstellar Medium, Xiangtan University, Xiangtan, Hunan 411105, China\\
\email{lvming@xtu.edu.cn}
\and
 Yunnan Observatories, Chinese Academy of Sciences (CAS), Kunming 650216, P.R. China
 \and
 Key Laboratory for the Structure and Evolution of Celestial Objects, CAS, Kunming 650216, P.R. China }
 

\abstract
{We combined a principal component analysis (PCA) and spectroscopy to investigate the origin of the soft excess in narrow-line Seyfert 1 galaxy Ark 564 with \textit{XMM-Newton} observations over a period of ten years. We find that the principal components in different epochs are very similar, suggesting stable variability patterns in this source. More importantly, although its spectra could be equally well fitted by the two soft excess models, simulations show that the principal components from the relativistically smeared reflection model match the data well. At the same time, the principal components from the warm corona model show significant inconsistency. This finding indicates that the soft excess in Ark 564 originates from the relativistically smeared reflection, rather than the Comptonization in the warm corona, thereby favoring the reflection origin or the "hybrid" origin of the soft excess. Furthermore, the presence of the narrow absorption features in the spectra suggests that the soft excess is unlikely to originate from absorptions due to possible outflowing winds. Our results indicate that the PCA coupled with spectral analysis is a promising approach to exploring the origin of the soft excess in active galactic nuclei (AGNs).}

{}
\keywords{X-rays}
\titlerunning{.}
\authorrunning{lyu et al.}
\maketitle

\section{Introduction}  

Soft X-ray excess appears as an excess in the soft band (< 2 keV) after the extrapolation of the hard X-ray continuum in active galactic nuclei (AGNs). After its first discovery \citep{singh85,arnaud85}, it has been widely observed in type 1 AGNs since there is no obscuration by the dusty torus \citep{Antonucci93}. Initially, soft X-ray excess was proposed to be the thermal radiation with the temperature of $\sim$ 0.1-0.2 keV originating from innermost part of the accretion flow. However, given that the disk temperature is inversely proportional to the black hole mass, the expected disk temperature for supermassive black holes should be typically of order 1-10 eV \citep[see, e.g.,][]{book}. Consequently, this scenario was disfavored by the fact that the standard accretion disk could not reach a temperature as high as 0.1 keV. It has been found that the temperature required for generating the soft excess is always $\sim$ 0.1-0.2 keV \citep[see, e.g.,][]{crummy06}, independent of the black hole mass.

At present, two popular, opposing models have been proposed to account for the soft excess: the warm corona model and the relativistically smeared reflection model. The former assumes that the soft excess comes from the up-scattered Comptonization of the ultraviolet (UV) seed disk photons in a warm (kT$\sim$ 0.1-1 keV) and optically thick ($\tau$$\sim$10-40) corona \citep[e.g.,][]{jin09,jin12,petrucci18}. This warm corona model is favored by the existence of similarities in the spectral shape and the variability between the optical/UV and the soft X-ray emission \citep[e.g.,][]{walter93,edelson96}. The relativistically smeared reflection model presumes that the low-energy emission-lines produced from the reflection off the inner part of the ionized disk are blurred and smoothed by strong relativistic effects around central black-hole, and finally forms the observed excess \citep[e.g.,][]{crummy06,jiang20}. This model involves the atomic transition process and hence could more easily explains why the soft X-ray excess remains consistent over a wide range of black hole mass. Besides, this scenario is favored by the results from the soft X-ray reverberation lags detection \citep{fabian09,demarco11,cackett13,demarco13,chiang17,mallick18,alston20}. In addition to the  two models above, it has also been proposed that the soft excess could be due to the relativistically blurred absorption by disk winds \citep[e.g.,][]{gierlinski04}. In this absorption scenario, the complex velocity structure of disk winds leads to substantial broadening, which masks the sharp absorption features and generates a smooth soft excess structure in the spectrum.

Observationally, the above two models could reproduce the observed X-ray spectrum well \citep[e.g.,][]{pal16,waddell19}, indicating that spectroscopy alone is not enough to distinguish them. Recent years, investigations into the Narrow-line Seyfert 1 galaxies (NLSy1s) and the broad-line Seyfert 1 galaxies (BLSy1s) have reported a variety of conclusions about the origin of the soft excess. In particular, NLSy1s are a subset of type-1 AGNs with the full-width-half-maximum (FWHM) of the H$\beta$ lines less than 2000 km/s, while BLSy1s have FWHM bigger than 2000 km/s \citep{Oster85,Goodrich89}. \citet{gliozzi20} studied the soft excess in 30 NLSy1s plus 59 BLSy1s and detected a positive correlation between the relative strength of the soft excess (R) and the primary X-ray spectral index. Besides, they found no correlation between the relative strength and the primary X-ray luminosity. These results indicate that the soft excess is generated by a warm corona. On the contrary, \citet{waddell20} found that the relative strength of the soft excess significantly correlates with the relative strength of hard excess for 22 NLSy1s, supporting the relativistically smeared reflection origin of the soft excess. Later, based on the Swift observations, \citet{ding22} found no clear correlation between the relative strength R and power-law spectra index, favoring the relativistically smeared reflection scenario.

Principal component analysis (PCA) is a powerful tool to dig different variability patterns in complex datasets. By decomposing a data set into several orthogonal eigenvectors, or principal components (PCs), PCA can efficiently quantify the variable components. The main advantage of the PCA is that it could obtain the variability in each separate varying component instead of only a total variability. When applied to a set of spectra, it returns detailed spectra of each variable component in a model-independent way. 

PCA has been widely used in many regions of astronomy, including stellar classifications \citep{whitney83}, variability in both AGNs \citep{mittaz90} and X-ray binaries \citep{malzac06,koljonen13}. \citet{parker14a} applied the PCA to study the spectral variability of MCG-06-30-15 and found that its variability is driven mainly by the variation of the power-law component normalization ($\sim$ 97\% variability), the photon index ($\sim$ 2\%), and the normalization of the reflection ($\sim$ 0.5\%). \citet{parker14b} indicated that variability component in NGC 1365 from the PCA is different from those in MCG-06-30-15. Furthermore, they found that the PCA provides a clear distinction between absorption and reflection as the driver of the variability in AGN spectra. \citet{parker15} applied the PCA to 26 AGNs and identified at least 12 different variation patterns, corresponding to several different physical mechanisms. Besides, the work of \citet{parker15} suggests that the PCA could be an extremely powerful tool for distinguishing different variability patterns in AGNs. Later, \citet{waddell19} applied the PCA to the source Mrk 478 and found that only the blurred reflection model could reproduce the overall shape of the the dominant PC (90\%), although its spectra could be nearly as equally well fitted by the partial covering model, the soft Comptonization model and the reflection model.
 \\
\\

Due to the degeneracy in the goodness of fits with the warm corona model and the blurred reflection model, the nature of the soft X-ray excess is still not fully understood after about 40 years since its first discovery. In this paper, we aim to combine the spectroscopy and the PCA to study the soft excess in the  NLSy1 AGN Arakelian 564 (Ark 564) with \textit{XMM-Newton} observations. Ark 564 locates at a redshift z = 0.02468 \citep{huchra99} and is one of the brightest  NLSy1s, with a 2-10 keV luminosity $\sim$ 2$\times10^{43}$ erg s$^{-1}$ \citep{turner01}. In the following analysis, we first fit the spectra with the two models separately and then simulated the PCs from each model based on the fitting results. Finally, we compared the simulated PCs with the one derived from real data to see their consistency.

\section{Observations and data reduction}
In this work, we analysed 13 \textit{\textit{XMM-Newton}} observations (Table 1) taken from 2000-06-17 to 2018-12-03 using the
European Photon Imaging Camera, EPIC-PN \citep{struder01} in imaging mode. We used the Science Analysis System (SAS) version 21.0.0 for the \textit{XMM-Newton} data reduction, with the latest calibration files applied. We run the tool {\tt epproc} for the extraction of the calibrated events, and used the command {\tt barycen} to convert the arrival time of photons from the local satellite frame to the barycenter of the solar system. We filtered the flaring particle background via an iterative process that leads to a maximization of the signal-to-noise ratio (S/N), the same as the method described in \citet{piconcelli04}. We selected only single and double events,  excluding all events at the edge of a CCD and close to a bad pixel for the extraction. We detected a moderate pile-up effect in some observations (Table 1) using the command {\tt epatplot}. To remove the pile-up effect, we finally selected events from circular regions of radius 40 arcsec region centred on the source position, excluding a central circular region of 10 arcsec and 8 arcsec for the Obs 11 and other pile-up observations, respectively. The background spectra was generated from a 50 arcsec circular region far from the source. We finally applied the command {\tt rmfgen} and {\tt arfgen} to produce the response matrices (RMFs) and ancillary response files (ARFs).

\begin{table*}
\centering
\tiny
\caption{\textit{XMM-Newton} observations of Ark 564 in this work. 
}

\begin{threeparttable}

\begin{tabular}{lcccc}
\hline
 ObsID   & Star time of exposure  & Pile-up  & Exposure (ks)\tnote{*} & Net count rate (cts/s)   \\
\hline

0006810101 (Obs 1)&  2000-06-17 12:08  & Yes    &  7.3  &  19.5  \\
0006810301 (Obs 2)& 2001-06-09 08:37   & No     &  7.1  &  23.7  \\
0206400101 (Obs 3)&  2005-01-05 19:45  & No     & 69.2  &  29.2  \\ 
0670130201 (Obs 4)& 2011-05-24 05:59    & Yes   & 40.9  &  22.6   \\
0670130301 (Obs 5)&  2011-05-30 15:14   & No    & 38.3  &  31.2   \\
0670130401 (Obs 6)&  2011-06-05 23:01   & No    & 37.9  &  31.4   \\
0670130501 (Obs 7)&  2011-06-11 17:04   & Yes   & 46.5  &  18.4   \\
0670130601 (Obs 8)&  2011-06-17 04:22   & No    & 40.3  &  33.4   \\
0670130701 (Obs 9)&  2011-06-25 23:23   & No    & 31.5  &  21.1   \\
0670130801 (Obs 10)&  2011-06-29 06:55   & Yes  & 40.5  &  15.0    \\
0670130901 (Obs 11)&  2011-07-01 06:53   & Yes  & 38.6  &  16.9    \\
0830540101 (Obs 12)&  2018-12-01 15:49   & No   & 76.6  &  24.5    \\
0830540201 (Obs 13)&  2018-12-03 15:51   & No   & 72.6  &  27.4    \\

\hline          

\end{tabular}
   
\begin{tablenotes}
\tiny
\item[*] Final PN exposure time after excluding instrumental drops and background flares.   
\end{tablenotes}   

\end{threeparttable}   
   
\medskip        
\label{obs}     
\end{table*}

\section{Principal component analysis}
We applied the PCA in the same way as the one described in the work of Parker et al. (2014). The steps could be briefly summarized as below: (1) we first divided each observation into 5 ks segments, and extracted spectra for each segment; (2) we calculated a mean background-subtracted spectrum of, F$_\textrm{mean}(E_\mathrm{j})$, over 0.4-9.0 keV for all segments; (3) we subtracted the mean spectrum from each segment spectrum to obtain a set of fractional residual spectra using the formula, $F_\textrm{res,i}(E_\textrm{j})=[F_\textrm{i}(E_\textrm{j})-F_\textrm{mean}(E_\mathrm{j})]/F_\textrm{mean}(E_\mathrm{j})$, where F$_\textrm{i}(E_\textrm{j})$ is the flux at the j-th energy bin in the i-th segment spectrum; (4) we applied the singular value decomposition (SVD) to the F$_\textrm{res,i}(E_\textrm{j})$ matrix to derive the orthogonal eigenvectors and eigenvalues. We obtained the fractional variability of each component by dividing each eigenvalue by the sum of all eigenvalues; and (5) we perturbed the observed photon counts by a random amount commensurate with the photon shot noise and then applied the PCA to this perturbed dataset to make an estimate of the variance on the component for the error calculation \citep{miller07}. Each eigenvector obtained in step (4) is the spectrum of an individually varying component and the eigenvalue is connected to the variation that component is responsible for.

\section{Spectral analysis}
In this work, we used XSPEC version 12.13.1 (Arnaud 1996) to fit all the 13 \textit{XMM-Newton} spectra together in 0.4-10.0 keV energy range. We used the model {\sc tbabs} to describe the absorption attributed to the interstellar medium (ISM) along the line of sight (LoS), with the solar abundance table of \citet{wilms00} and the photoionization cross-section table of \citet{verner96}. The column density was fixed at $N_{H}=5.34 \times 10^{20}$ $cm^{-2}$ \citep{kalberla05} for all the fits.

\subsection{Warm corona model}
We applied the model {\sc nthcomp} \citep{zdziarski96,zycki99} to describe the power law emission from the hot corona. The thermal Comptonized model {\sc nthcomp} more accurately describes  the high-energy shape and the low-energy rollover compared with an exponentially cut-off power-law component. For the soft excess emission, we first used a simple {\sc bbody} model to estimate its shape. And we finally replaced the {\sc bbody} model with a physical model {\sc comptt} \citep{titar94} to account for the emission from the possible warm corona. Overall, {\sc comptt} is an analytic model describing Comptonization of soft photons in a plasma. It works well for both the optically thin and thick regimes. During the fitting process, the redshift to the source, z, is fixed at 0.02468 \citep{huchra99} and the disk seed photon temperature $kT_{seed}$ is frozen at 0.05 keV \citep[e.g.,][]{sarma15} in both the {\sc nthcomp} and {\sc comptt} component. We fixed the electron temperature, $kT_{e}$, in the {\sc nthcomp} model at 100 keV and selected the disk geometry in the {\sc comptt}.  

 \subsection{Relativistically smeared reflection model}
For the reflection scenario, we selected the model {\sc relxillcp}, which describes the reflection off the disc illuminated by a power-law source. The {\sc relxillcp}\footnote{http://www.sternwarte.uni-erlangen.de/~dauser/research/relxill/} combines the {\sc xillver} model \citep{garcia13} and the {\sc relline} code \citep{dauser10}, and it calculates the reflection from each emission angle. The parameters are: the inclination angle, $i$; the spin parameter, $a$; the inner and outer radius of the disk, $R_{in}$ and $R_{out}$; the breaking radius separating them, $R_{break}$; the emissivity index of the inner and the outer disk, $q_{in}$ and $q_{out}$; the redshirt, $z$; the photon index, $\Gamma$; the density and the ionization parameter of the disk, $N;$ and $\xi$, the iron abundance, the electron temperature, $kT_{e}$; the reflection fraction, $R_{refl}$; and the normalization. During the fit, we fixed the spin at 0.998, which is the same as the value in \citet{lewin22}. The inner radius, $R_{in}$, is frozen at 1 $R_{ISCO}$, while the outer radius and the breaking radius are fixed at 400 $R_{g}$. The two emissivities are linked to be the same $q_{in}$=$q_{out}$. We set the electron temperature, $kT_{e}$, at 100 keV and fixed the iron abundance to the solar abundance. The inclination angle between different observations are linked together and we do the same to the emissivity index.
\\
\\
After applying the above warm corona model and the reflection model, we found that there are still some residuals in the fits, possibly due to the complex absorption in Ark 564. We finally found that these residuals could be well described by a model composed of a partial ionized absorber ({\sc zxipcf}), along with three to four phenomenological low-energy absorption edges ({\sc zedge}) in the soft energy band, similarly to the absorption model used in \citet{sarma15}.

\section{PC simulations}
With the parameters derived from the above spectral analysis, we  constructed a set of fake spectra and generated the simulated PCs. The fake spectra were produced by the tool {\sc Fakeit} in Xspec. We varied the free parameters randomly within the ranges derived from the fits: $\Gamma$, $kT_{e}$, $\tau$, $Nor_{nthcomp}$, and  $Nor_{compTT}$ for the warm corona model; $\Gamma$, log($\xi$), log(N), $R_{refl}$, and $Nor_{relxillcp}$ for the reflection model. For the free parameters in the absorption model, we first fixed them at their mean value, which we calculated from the values in the joint fitting results. In addition to that, we also let them vary randomly in their variation ranges in the related simulations to see the possible influence of the absorption in the PCs. We generated 200 simulated fake spectra for each model, with the exposure of each one being 5 ks, which is the same as the length of the segments in the previous principle component analysis. Finally, we applied the PCA to the fake spectra and obtained the corresponding PCs.

\begin{sidewaystable}  
\setlength{\tabcolsep}{2pt}
\tiny
\centering
\caption{Best-fitting results for the joint fit to the \textit{XMM-Newton} spectra of Ark 564 with the warm corona model Tbabs $\times$ (Nthcomp+CompTT). The chi-square of the fit is $\chi^2_\nu$ ($\chi^2/dof)$=1.14 (2239/1960). All errors in the Tables are at the 90 percent confidence level unless otherwise indicated. A symbol $*$ means that the error pegged at the hard limit of the parameter range.}
\begin{tabular}{ccccccccccccccc}
\hline
Model Comp   &   Parameter     & Obs 1       & Obs 2       & Obs 3       & Obs 4       & Obs 5        & Obs 6  &  Obs 7  & Obs 8       & Obs 9       & Obs 10       & Obs 11       & Obs 12        & Obs 13   \\
\hline
 zxipcf    & N$_{H}$ (10$^{22}$)  &    24$_{-12}^{+101}$        &  162$_{-24}^{+103}$          &  49$_{-12}^{+29}$         &  142$\pm 15$                &  65$\pm 14$               &  23$_{-3}^{+7}$           & 23$_{-5}^{+7}$           &  45$_{-7}^{+12}$        & 64$_{-14}^{+20}$           & 67$_{-16}^{+34}$        & 65$_{-20}^{+38}$            &   57$_{-23}^{+8}$         & 24$_{-2}^{+7}$            \\   
           & log($\xi$)      &         0.01$_{-0*}^{+5.17}$      &  1.7$\pm 0.7$                &  1.6$\pm 0.5$             &  2.1$_{-0.1}^{+0.2}$        &  1.9$_{-0.5}^{+0.1}$      &  1.1$_{-0.8}^{+0.4}$      & 0.9$_{-0.8}^{+0.5}$      &  0.9$\pm 0.6$           & 1.9$_{-1.0}^{+0.1}$        & 2.2$\pm 0.2$             & 2.1$_{-0.1}^{+0.2}$         &   1.9$_{-0.9}^{+0.1}$     & 0.01$_{-0*}^{+0.35}$      \\   
           & $f_{cover}$     &         0.23$_{-0.12}^{+0.52}$    &  0.79$_{-0.12}^{+0.09}$      &  0.31$\pm 0.04$           &  0.65$\pm 0.03$             &  0.37$\pm 0.04$           &  0.26$\pm 0.03$           & 0.27$\pm 0.04$           &  0.38$\pm 0.07$         & 0.41$\pm 0.05$             & 0.38$\pm 0.06$           & 0.31$_{-0.04}^{+0.02}$      &   0.41$\pm 0.04$          & 0.34$_{-0.03}^{+0.13}$    \\   
 zedge1     & $E_{c}$ (keV)  &         0.554$\pm 0.015$          &  0.530$\pm 0.012$            &  0.526$\pm 0.004$         &  0.543$\pm 0.009$           &  0.549$\pm 0.011$         &  0.540$\pm 0.007$         & 0.548$\pm 0.007$         &  0.533$\pm 0.006$       & 0.547$\pm 0.009$           & 0.532$\pm 0.008$         & 0.542$\pm 0.008$            &   0.526$\pm 0.005$        & 0.525$\pm 0.004$          \\   
           & $\tau$          &         0.19$\pm 0.05$            &  0.20$\pm 0.05$              &  0.20$\pm 0.01$           &  0.11$\pm 0.02$             &  0.11$\pm 0.02$           &  0.14$\pm 0.02$           & 0.16$\pm 0.02$           &  0.11$\pm 0.02$         & 0.16$\pm 0.02$             & 0.19$\pm 0.03$           & 0.14$\pm 0.02$              &   0.16$\pm 0.02$          & 0.14$\pm 0.02$            \\   
 zedge2     & $E_{c}$ (keV)  &         0.727$\pm 0.019$          &  0.719$\pm 0.015$            &  0.699$\pm 0.005$         &  0.714$\pm 0.017$           &  0.708$\pm 0.012$         &  0.714$\pm 0.007$         & 0.724$\pm 0.008$         &  0.715$\pm 0.009$       & 0.702$\pm 0.010$           & 0.687$\pm 0.014$         & 0.704$_{-0.007}^{+0.013}$   &   0.707$\pm 0.007$        & 0.707$\pm 0.006$          \\   
           & $\tau$          &         0.15$\pm 0.04$            &  0.16$\pm 0.05$              &  0.14$\pm 0.01$           &  0.07$\pm 0.02$             &  0.11$\pm 0.02$           &  0.14$\pm 0.02$           & 0.14$\pm 0.02$           &  0.10$\pm 0.02$         & 0.16$\pm 0.02$             & 0.13$\pm 0.02$           & 0.10$\pm 0.02$              &   0.14$\pm 0.01$          & 0.12$\pm 0.01$            \\   
 zedge3     & $E_{c}$ (keV)  &         1.17$_{-0.10}^{+0.13*}$   &  1.0$_{-0*}^{+0.3*}$         &  1.14$\pm 0.03$           &  1.20$\pm 0.03$             &  1.21$\pm 0.03$           &  1.17$\pm 0.09$           & 1.19$\pm 0.08$           &  1.13$\pm 0.03$         & 1.14$\pm 0.05$             & 1.11$\pm 0.02$           & 1.16$_{-0.12}^{+0.08}$      &   1.17$\pm 0.02$          & 1.18$\pm 0.02$            \\   
           & $\tau$          &         0.034$_{-0.032}^{+0.040}$ &  0.029$_{-0.028*}^{+0.041}$  &  0.038$\pm 0.010$         &  0.043$_{-0.013}^{+0.007}$  &  0.045$\pm 0.013$         &  0.017$\pm 0.012$         & 0.022$\pm 0.014$         &  0.044$\pm 0.012$       & 0.048$\pm 0.016$           & 0.082$\pm 0.016$         & 0.021$\pm 0.016$            &   0.068$\pm 0.010$        & 0.070$\pm 0.010$          \\   
 \\
 nthComp   & $\Gamma$        &         2.52$\pm 0.07$            &  2.52$\pm 0.06$              &  2.58$\pm 0.02$           &  2.64$\pm 0.02$             &  2.62$\pm 0.03$           &  2.59$\pm 0.03$           & 2.59$\pm 0.03$           &  2.60$\pm 0.03$         & 2.57$\pm 0.04$             & 2.56$\pm 0.03$           & 2.62$\pm 0.03$              &   2.59$\pm 0.03$          & 2.61$\pm 0.02$            \\   
           & Norm (10$^{-2}$)&         2.45$_{-0.61}^{+1.04}$    &  4.75$_{-1.23}^{+4.40}$      &  2.12$_{-0.17}^{+0.12}$   &  6.74$_{-0.74}^{+0.89}$     &  2.27$_{-0.16}^{+0.71}$   &  2.04$_{-0.10}^{+0.21}$   & 2.47$_{-0.16}^{+0.21}$   &  2.60$_{-0.30}^{+0.24}$ & 1.64$_{-0.16}^{+0.25}$     & 2.39$_{-0.27}^{+0.24}$   & 3.05$_{-0.37}^{+0.13}$      &   1.97$_{-0.15}^{+0.18}$  & 1.92$_{-0.11}^{+0.40}$    \\   
 compTT    & $kT_{e}$ (keV)  &         0.157$\pm 0.014$          &  0.160$_{-0.017}^{+0.030}$   &  0.145$\pm 0.004$         &  0.153$\pm 0.006$           &  0.152$\pm 0.004$         &  0.143$\pm 0.002$         & 0.144$\pm 0.002$         &  0.152$\pm 0.007$       & 0.149$_{-0.003}^{+0.006}$  & 0.152$\pm 0.007$         & 0.148$\pm 0.006$            &   0.150$\pm 0.004$        & 0.152$\pm 0.004$          \\   
           & $\tau$          &         40$_{-10}^{+24}$          &  28$_{-5}^{+9}$              &  46$_{-4}^{+10}$          &  100$_{-40}^{+0*}$          &  56$_{-10}^{+27}$         &  100$_{-38}^{+0*}$        & 100$_{-40}^{+0*}$        &  40$_{-5}^{+10}$        & 100$_{-41}^{+0*}$          & 100$_{-42}^{+0*}$        & 100$_{-45}^{+0*}$           &   55$_{-9}^{+21}$         & 46$_{-5}^{+10}$           \\   
           & Norm            &         1.09$_{-0.22}^{+0.85}$    &  4.71$_{-2.01}^{+3.26}$     &  0.90$_{-0.15}^{+0.22}$   &  0.99$_{-0.07}^{+0.40}$     &  0.65$\pm 0.22$           &  0.43$_{-0.03}^{+0.10}$   & 0.61$_{-0.04}^{+0.13}$   &  0.92$\pm 0.19$         & 0.39$_{-0.03}^{+0.28}$     & 0.48$_{-0.06}^{+0.23}$   & 0.60$_{-0.05}^{+0.26}$      &   0.60$_{-0.07}^{+0.10}$  & 0.71$\pm 0.13$            \\   
\hline                                                                                                                    
\end{tabular}                                                                                                                                                                                                                    

\medskip  
\label{warmcorona}
\end{sidewaystable}


\begin{sidewaystable}
\setlength{\tabcolsep}{2pt}
\tiny
\caption{Best-fitting results for the joint fit to the \textit{XMM-Newton} spectra of Ark 564 with the reflection model Tbabs $\times$ RelxillCp. The chi-square of the fit is $\chi^2_\nu$ ($\chi^2/dof)$=1.17 (2256/1932).}
\begin{tabular}{ccccccccccccccc}
\hline
Model Comp   &   Parameter     & Obs 1       & Obs 2       & Obs 3       & Obs 4       & Obs 5        & Obs 6  &  Obs 7  & Obs 8       & Obs 9       & Obs 10       & Obs 11       & Obs 12        & Obs 13   \\
\hline
 zxipcf    & N$_{H}$ (10$^{22}$)&      288$_{-283}^{+116}$       &   84$_{-31}^{+119}$         &  181$_{-25}^{+39}$         &  83$_{-16}^{+20}$         &  181$_{-58}^{+34}$           & 18$_{-3}^{+10}$              &    17$_{-6}^{+34}$               &   65$_{-7}^{+25}$              &   18$_{-3}^{+18}$          &    46$_{-9}^{+11}$         &     15$\pm 4$                  &    58$_{-9}^{+4}$             &     65$_{-9}^{+20}$           \\          
           & log($\xi$)      &         4.39$_{-0.89}^{+1.48}$   &   0.76$_{-0.75}^{+1.49}$    &  2.71$_{-0.15}^{+0.05}$    &  2.03$\pm 0.07$           &  2.73$_{-0.17}^{+0.05}$      & 1.14$_{-0.72}^{+0.31}$       &    0.96$_{-0.95*}^{+0.73}$       &    1.09$_{-0.41}^{+0.53}$       &   0.01$_{-0*}^{+1.02}$     &    1.99$\pm 0.06$          &     0.83$_{-0.82*}^{+0.58}$    &    1.91$_{-0.42}^{+0.01}$     &     1.33$_{-0.55}^{+0.50}$    \\          
           & $f_{cover}$     &         0.50$_{-0.41}^{+0.02}$    &   0.42$\pm 0.07$            &  0.36$\pm 0.03$            &  0.38$\pm 0.03$           &  0.40$\pm 0.04$              & 0.26$\pm 0.02$               &    0.20$\pm 0.03$                &    0.32$_{-0.03}^{+0.08}$       &   0.35$\pm 0.03$           &    0.32$\pm 0.05$          &     0.35$\pm 0.03$             &    0.35$\pm 0.02$             &     0.35$\pm 0.02$            \\          
 zedge1    & $E_{c}$ (keV)   &         0.39$\pm 0.02$            &   0.40$_{-0.08}^{+0.02}$    &  0.35$\pm 0.03$            &  0.39$\pm 0.02$           &  0.396$\pm 0.005$            & 0.38$\pm 0.03$               &    0.32$\pm 0.02$                &    0.37$\pm 0.02$               &   0.38$\pm 0.02$           &    0.39$_{-0.04}^{+0.01}$  &     0.33$\pm 0.03$             &    0.392$\pm0.006$            &     0.403$\pm 0.005$          \\          
           & $\tau$          &         1.00$_{-0.08}^{+0.13}$    &   0.88$\pm 0.14$            &  0.96$_{-0.13}^{+0.20}$    &  1.00$\pm 0.09$           &  1.12$\pm 0.05$              & 0.96$_{-0.05}^{+0.18}$       &    1.65$_{-0.26}^{+0.37}$        &    0.84$_{-0.07}^{+0.12}$       &   1.24$_{-0.11}^{+0.19}$   &    0.89$_{-0.09}^{+0.14}$  &     1.61$\pm 0.40$             &    1.01$\pm 0.05$             &     1.09$\pm 0.04$            \\          
 zedge2    & $E_{c}$ (keV)   &         0.53$\pm 0.01$            &   0.52$\pm 0.01$            &  0.513$\pm 0.003$          &  0.524$\pm 0.006$         &  0.530$\pm 0.005$            & 0.519$\pm 0.004$             &    0.518$\pm 0.004$              &    0.513$\pm 0.004$             &   0.524$\pm 0.004$         &    0.517$\pm 0.006$        &     0.514$\pm 0.004$           &    0.517$\pm 0.004$           &     0.525$\pm 0.004$          \\          
           & $\tau$          &         0.29$\pm 0.05$            &   0.34$\pm 0.05$            &  0.30$\pm 0.01$            &  0.28$\pm 0.03$           &  0.28$\pm 0.02$              & 0.26$\pm 0.02$               &    0.29$_{0.02}^{+0.05}$         &    0.25$\pm 0.02$               &   0.32$\pm 0.02$           &    0.29$\pm 0.04$          &     0.32$\pm 0.04$             &    0.31$\pm 0.02$             &     0.29$\pm 0.02$            \\          
 zedge3    & $E_{c}$ (keV)   &         1.14$\pm 0.05$            &   1.02$_{-0.02*}^{+0.07}$   &  1.07$\pm 0.02$            &  1.05$\pm 0.03$           &  1.04$\pm 0.02$              & 1.10$\pm 0.03$               &    1.09$\pm 0.04$                &    1.08$\pm 0.02$               &   1.11$\pm 0.03$           &    1.08$\pm 0.02$          &     1.08$\pm 0.04$             &    1.12$\pm 0.02$             &     1.08$\pm 0.02$            \\          
           & $\tau$          &         0.076$\pm 0.031$          &   0.067$\pm 0.029$          &  0.066$\pm 0.013$          &  0.051$\pm 0.015$         &  0.047$\pm 0.013$            & 0.058$_{-0.019}^{+0.012}$    &    0.048$\pm 0.018$              &    0.065$\pm 0.015$             &   0.074$\pm 0.022$         &    0.109$\pm 0.018$        &     0.063$\pm 0.017$           &    0.091$_{-0.007}^{+0.013}$  &     0.071$\pm 0.011$          \\          
 zedge4    & $E_{c}$ (keV)   &         1.36$\pm 0.07$            &   1.23$\pm 0.06$            &  1.24$\pm 0.03$            &  1.23$\pm 0.03$           &  1.24$\pm 0.02$              & 1.33$_{-0.09}^{+0.05}$       &    1.25$\pm 0.04$                &    1.24$\pm 0.03$               &   1.29$\pm 0.05$           &    1.27$\pm 0.03$          &     1.34$_{-0.09}^{+0.04}$     &    1.30$\pm 0.04$             &     1.24$\pm 0.02$            \\          
           & $\tau$          &         0.07$\pm 0.04$            &   0.08$\pm 0.04$            &  0.06$\pm 0.01$            &  0.07$\pm 0.02$           &  0.08$\pm 0.01$              & 0.04$\pm 0.01$               &    0.06$\pm 0.02$                &    0.05$\pm 0.01$               &   0.07$\pm 0.02$           &    0.08$\pm 0.02$          &     0.06$\pm 0.02$             &    0.06$\pm 0.01$             &     0.08$\pm 0.01$            \\          
 \\
 RelxillCp & $i$ (degree)    &                                   &                             &                            &                           &                              &                              &       52.6$\pm 0.5$              &                                 &                            &                            &                                &                               &                               \\          
           & Emissivity      &                                   &                             &                            &                           &                              &                              &       8.58$\pm 0.16$             &                                 &                            &                            &                                &                               &                               \\                                                                                                       
           & $\Gamma$        &         2.59$\pm 0.01$            &   2.65$\pm 0.03$            &  2.597$\pm 0.004$          &  2.59$\pm 0.01$           &  2.633$\pm 0.006$            & 2.68$\pm 0.01$               &    2.69$\pm 0.02$                &    2.639$_{-0.006}^{+0.009}$    &   2.72$\pm 0.02$           &    2.59$\pm 0.01$          &     2.82$\pm 0.02$             &    2.652$\pm 0.009$           &     2.644$\pm 0.006$          \\                                                                                                       
           & log($\xi$)      &         2.55$_{-0.16}^{+0.09}$    &   2.59$_{-0.09}^{+0.12}$    &  2.70$_{-0.02}^{+0.01}$    &  2.79$\pm 0.05$           &  2.70$_{-0.02}^{+0.01}$      & 2.70$_{-0.06}^{+0.02}$       &    2.68$_{-0.06}^{+0.05}$        &    2.70$_{-0.05}^{+0.04}$       &   2.71$_{-0.05}^{+0.03}$   &    2.67$\pm 0.07$          &     2.70$_{-0.06}^{+0.05}$     &    2.70$\pm 0.02$             &     2.70$_{-0.02}^{+0.01}$    \\                                                                                                       
           & log(N)          &         17.01$_{-0.43}^{+0.15}$   &   16.85$_{-0.32}^{+0.21}$   &  16.99$_{-0.07}^{+0.05}$   &  17.17$_{-0.28}^{+0.07}$  &  17.05$_{-0.04}^{+0.03}$     & 17.00$_{-0.17}^{+0.05}$      &    16.57$_{-0.16}^{+0.12}$       &    16.65$_{-0.11}^{+0.14}$      &   16.99$_{-0.17}^{+0.08}$  &    17.24$_{-0.21}^{+0.10}$ &     16.66$_{-0.19}^{+0.17}$    &    17.00$_{-0.09}^{+0.02}$    &     16.97$_{-0.12}^{+0.05}$   \\                                                                                                       
           & $R_{refl}$      &         11.27$_{-1.95}^{+2.03}$   &   12.96$_{-2.25}^{+2.46}$   &  7.88$_{-0.39}^{+0.29}$    &  12.09$_{-0.71}^{+0.79}$  &  11.37$_{-0.57}^{+0.68}$     & 4.12$_{-2.08}^{+0.36}$       &    7.27$\pm 0.64$                &    6.40$\pm 0.47$               &   5.08$_{-0.49}^{+0.53}$   &    7.98$\pm 0.68$          &     3.25$\pm 0.43$             &    6.78$_{-0.39}^{+0.44}$     &     10.27$_{-0.49}^{+0.64}$   \\                                                                                                       
           & Norm (10$^{-4}$)&         1.18$\pm 0.17$            &   1.11$_{-0.09}^{+0.17}$    &  1.46$\pm 0.04$            &  1.41$_{-0.07}^{+0.32}$   &  1.24$\pm 0.07$              & 1.86$_{-0.08}^{+0.12}$       &    1.84$_{-0.07}^{+0.11}$        &    1.81$\pm 0.08$               &   1.37$\pm 0.09$           &    1.24$_{-0.07}^{+0.10}$  &     4.02$\pm 0.24$             &    1.34$\pm 0.05$             &     1.17$\pm 0.05$            \\                                                                                                       
\hline                                
\end{tabular}                                                                                                                                                                                                                                                          

\medskip  
\label{reflection}
\end{sidewaystable}

\section{Results}
\subsection{PCA results}

\begin{figure}
\centering
\includegraphics[width=0.38\textwidth]{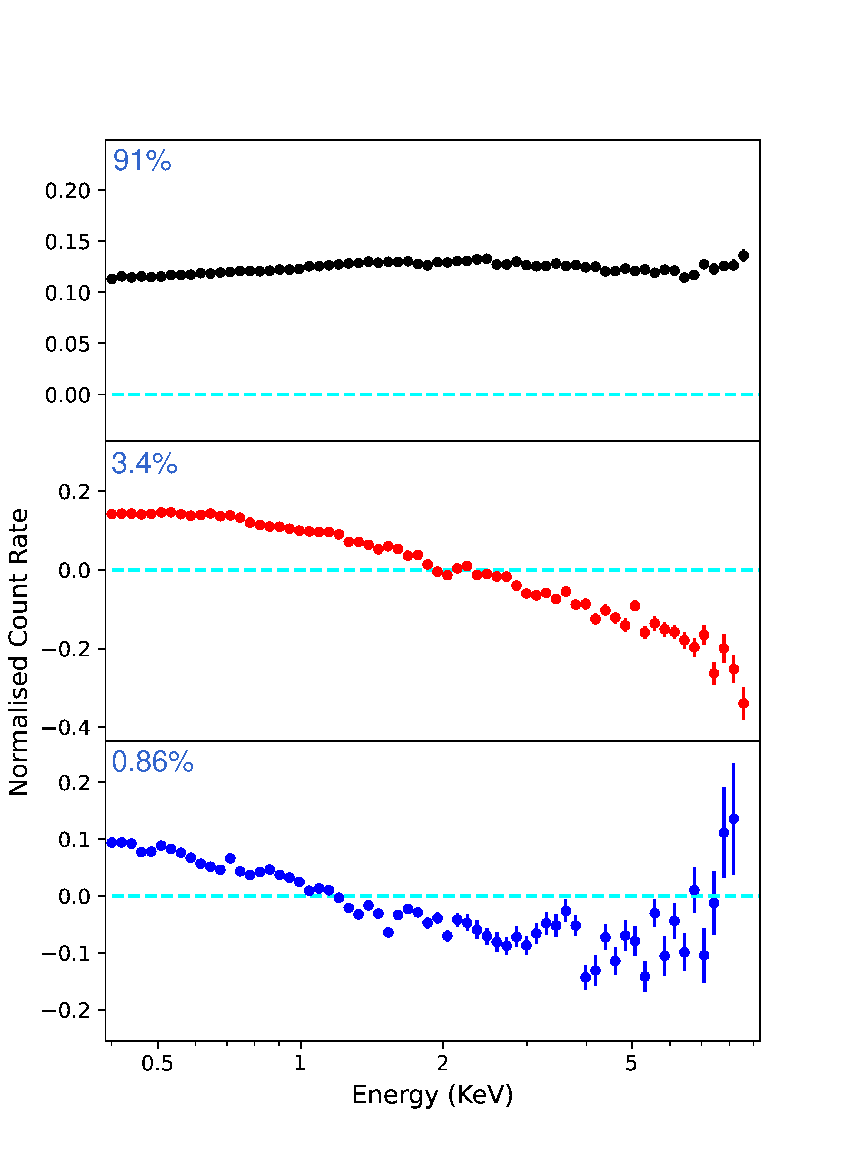}
\caption{Normalized spectra of the first three significant components (top to bottom) of the Ark 564 between 0.4-9.0 keV. We used the black, red and blue to mark the first, second, and third principle component, respectively. We binned the spectra logarithmically into 65 energy bins and calculated the errors by perturbing the input spectra. The zero level in each panel is marked with a green dashed line. We also show the variability percentage of the component in the top left corner of each panel. For the third component, we do not include the rightmost datapoint (x=8.6, y=0.84)  to show the detailed structure in the soft energy clearly.}
\label{pca}
\end{figure}

\begin{figure}
\centering
\includegraphics[width=0.41\textwidth]{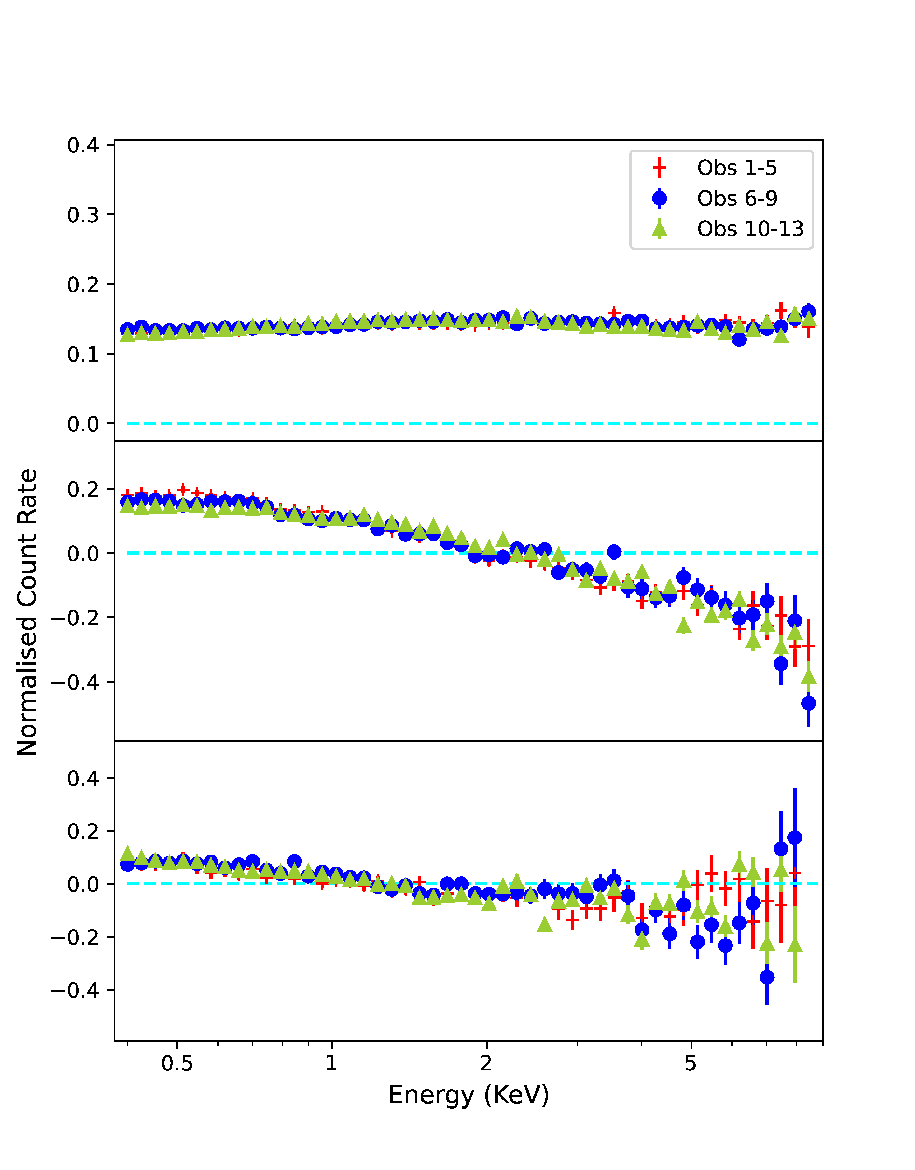}
\caption{First three significant PCs of Ark 564 in different epochs: Obs 1-5 (red cross symbol), Obs 6-9 (blue circle), and Obs 10-13 (green triangle). The variability percentage of the three components is 88\%-6\%-1\% for the Obs 1-5, 94\%-3\%-1\% for the Obs 6-9, and 92\%-3\%-1\% for the Obs 10-13, respectively.}
\label{different}
\end{figure}

In Fig \ref{pca}, we show the first three significant principal components based on all 13 \textit{XMM-Newton} observations. We found that the first component is always positive and remains relatively flat, showing suppression at low energies and at the energy of the iron line. The second (pivoting) component is positive below $\sim$ 2 keV and then becomes negative above that. Since values below zero indicate that the variation in the corresponding energy bins is anti-correlated with the positive bins, those points at low energies in this component significantly anti-correlated with the ones at high energies. Moreover, it is flattening at low energies, and then steepening at high energies. For the third component, it is positive at both the low-energy and the high-energy ends, while it is below zero at the energy in between, with a turnover at $\sim$ 4-7 keV. To see whether the principal components in Ark 564 vary with time, we further applied the PCA to datasets in different epochs composed of Obs 1-5, Obs 6-9, and Obs 10-13, respectively. In Fig \ref{different}, we show the PCs of Ark 564 in different epochs. We found that the shape of the first three significant PCs in different epochs are very similar, suggesting a consistent underlying physical mechanism driving the variability.

\subsection{Spectral analysis results}



\begin{figure}
\centering
\includegraphics[width=0.30\textwidth, angle=-90]{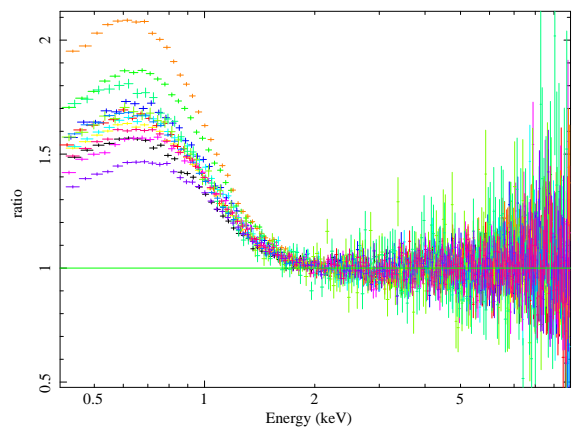}
\caption{Data versus model ratio plot obtained from the joint fit of the 13 \textit{XMM-Newton} spectra of Ark 564 with the phenomenological warm corona model {\sc tbabs $\times$ (bbody+nthcomp)}. The {\sc bbody} component was removed from the best fitting model in order to show the shape of the soft excess.}
\label{ratio}
\end{figure}

\begin{figure*}
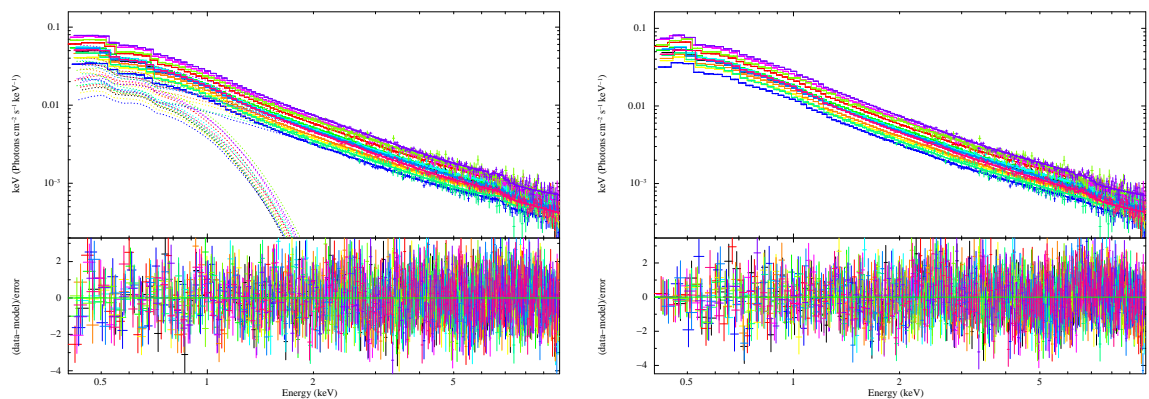

\center
\includegraphics[width=0.28\textwidth, angle=-90]{wc_fit.eps}
\includegraphics[width=0.28\textwidth, angle=-90]{ref_fit.eps}
\caption{Fitting results with the warm corona model (left) and the reflection model (right) for the \textit{XMM-Newton} observations of Ark 564. We show the fitted spectra and the individual model components (main panel), and the residuals in terms of sigmas (sub panel).}
\label{fit}
\end{figure*}

In Fig \ref{ratio}, we show the ratio between the data and the model after we removed the {\sc bbody} component from the phenomenological warm corona model. As shown in the plot, the soft excess below $\sim$ 2 keV is significant in all observations. Furthermore, there is a clear variation in the shape of the excess in these observations.

In Table \ref{warmcorona}, we show the parameters of the fit with the warm corona model {\sc comptt}. The $\Gamma$ values and the temperature of the warm corona, $kT_{e}$, remain more or less constant,  namely:\ $\sim$ 2.6 and $\sim$ 0.147 keV, in all cases. The normalization of the {\sc comptt} component distributes in a range from 0.36  to 7.97 in these observations. For the reflection model (Table \ref{reflection}), the photon index $\Gamma$ is from 2.58 to 2.84, with the reflection fraction in a wide range of $\sim$ 2-15. The inclination angle derived in the fit is 52.6$\pm$0.5 degrees, and the emissivity index is $\sim$ 8.6. For the accretion disk, it was highly ionised, with the value of the log($\xi$) ranges from 2.39 to 2.84. The disk density log(N) shows only small change, from 16.57$_{-0.16}^{+0.12}$ in Obs 7 to 17.24$_{-0.21}^{+0.10}$ in Obs 10. The corresponding spectra, individual components, and residuals of the fits are shown in Fig \ref{fit}. Apparently, both the warm corona model and the reflection model could well describe the observed spectra in the figure, with the main difference being that a separated soft component is present in order to describe the data in the warm corona model.

\subsection{Simulation results}
\subsubsection{Warm corona model}

\begin{figure*}
\center
\includegraphics[width=0.35\textwidth]{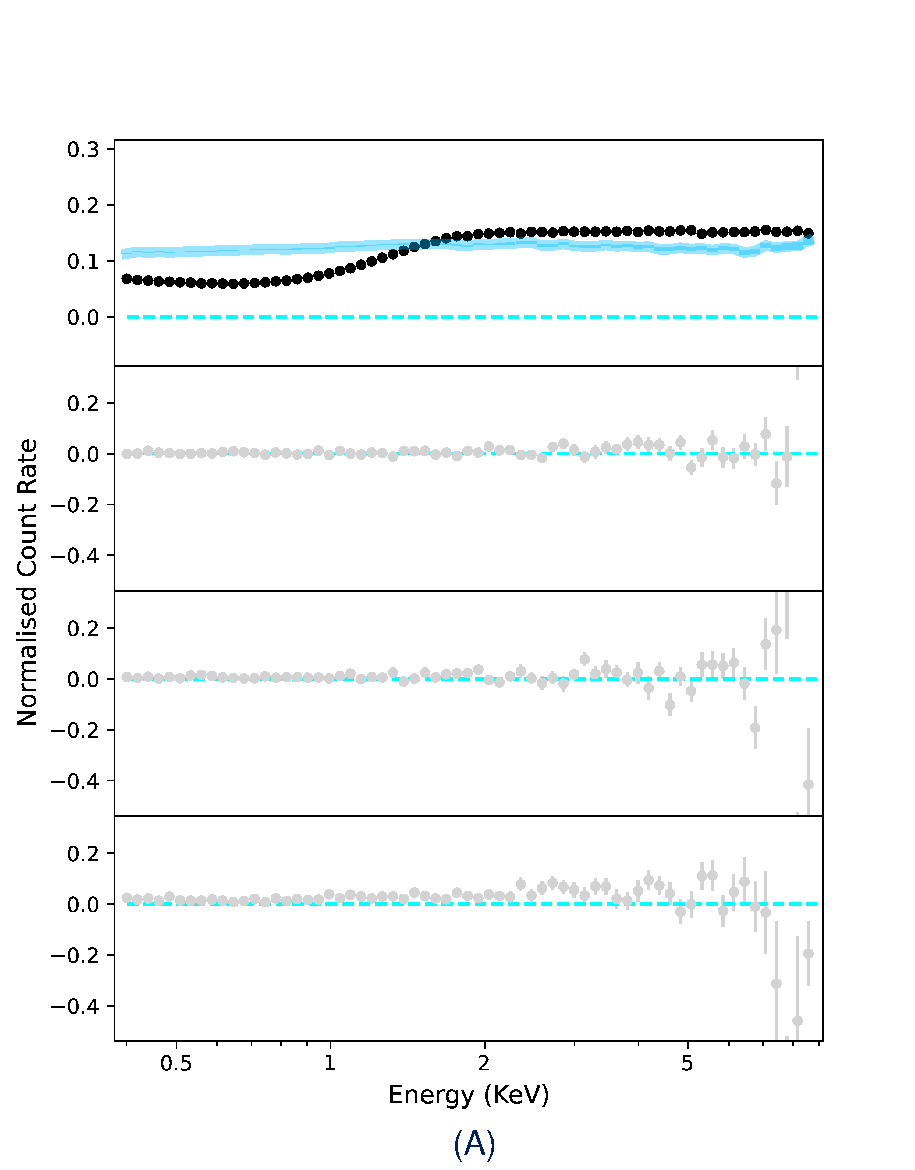}
\includegraphics[width=0.35\textwidth]{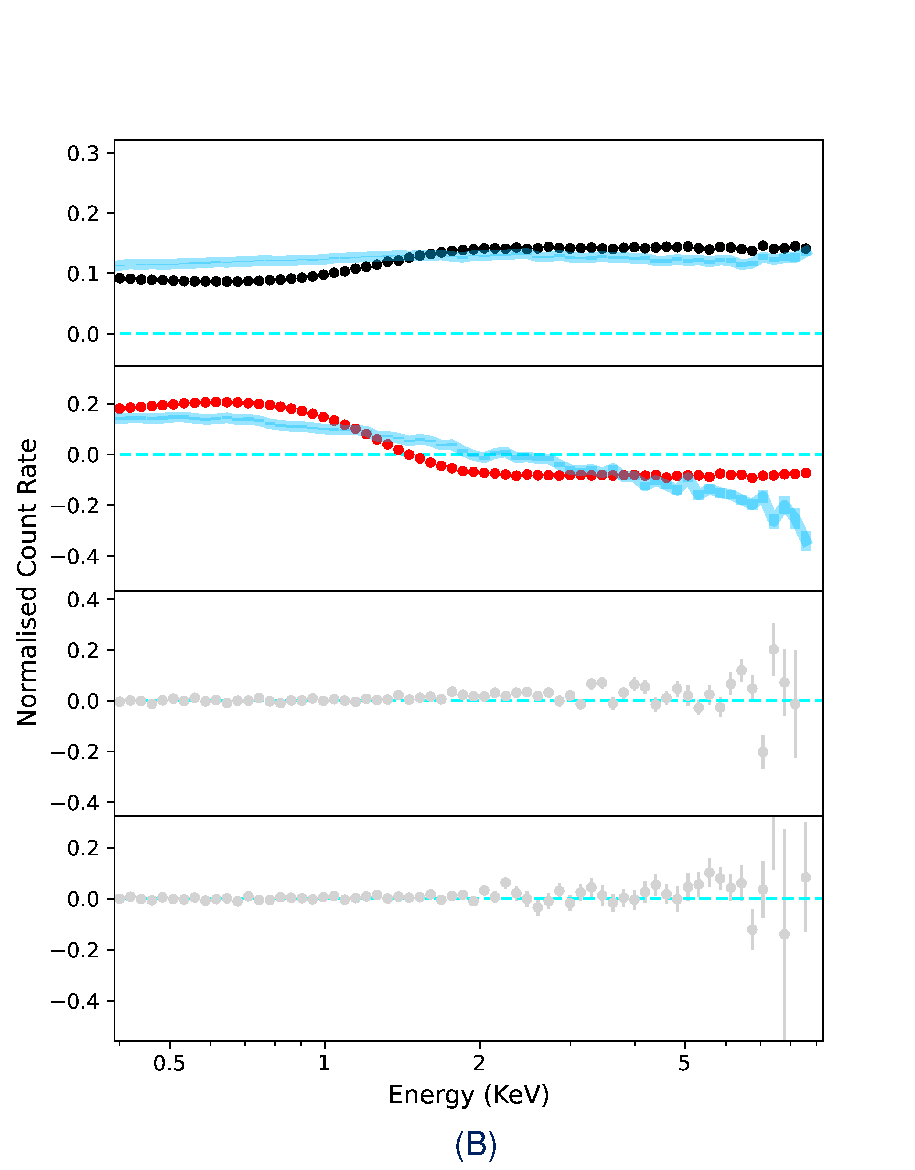}
\includegraphics[width=0.35\textwidth]{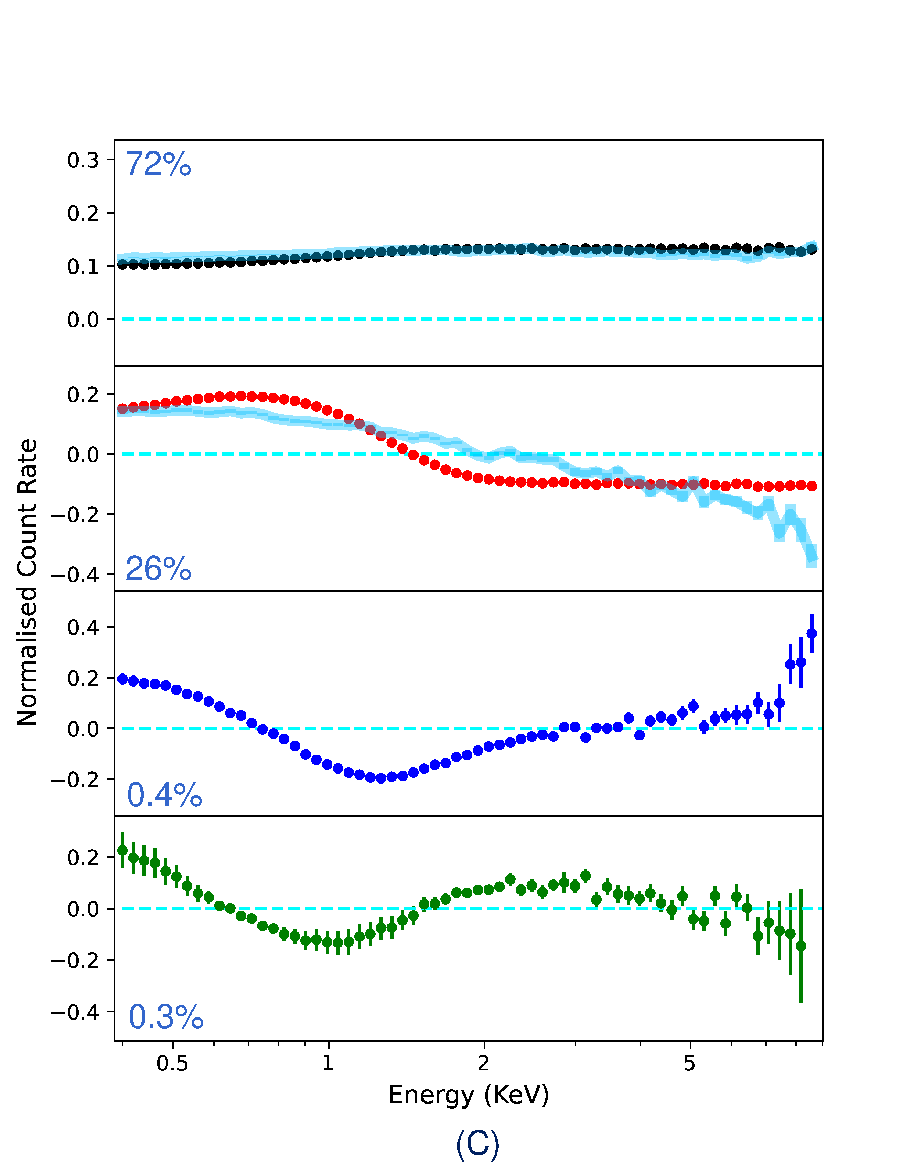}
\includegraphics[width=0.35\textwidth]{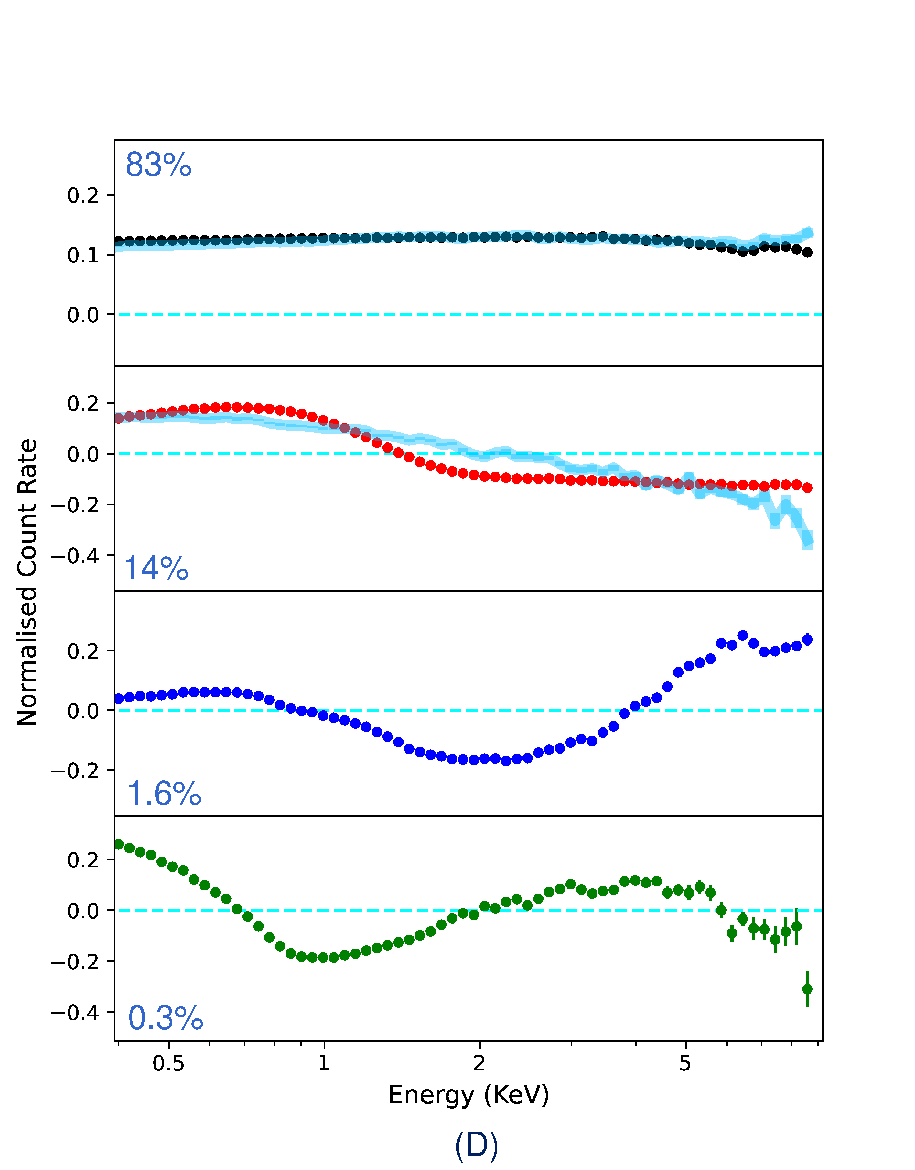}

\caption{Simulated PCs from the variations of different parameters in the warm corona model. We first let the parameter Nor$_{nth}$ vary randomly and there appears a principle component above the zero-line (panel A). When we further vary the parameter, Nor$_{comptt}$,  the second component appears (panel B). Next, we set all the free parameters in the continuum {\sc nthcomp} and {\sc comptt} free to change (panel C). Finally, we vary the free parameters in the absorption model {\sc zxipcf} and {\sc zedge} to see the possible influence of the absorption effect (panel D). In the figure, we also plotted the first and the second PC from real data (cyan curves) for the purpose of comparison. }
\label{simu_wc}
\end{figure*}

We show the simulated PCs from the warm corona model in Figure \ref{simu_wc}. In panel A, a component always above zero is present when we let the parameter Nor$_{nth}$ vary randomly. This PC is suppressed below $\sim$ 2 keV and remains flat in most of the energies. In panel B, we show the PCs when another normalisation Nor$_{comptt}$ also varies. In this case there appears a new pivoting component, whose overall shape is relatively flat above $\sim$ 2 keV. While the first component, compared with the one in panel A, becomes less suppressed below $\sim$ 1 keV. 

In panel C, we show the PCs when all the free parameters in the continuum {\sc nthcomp} and {\sc comptt} vary. We found that the first component is relative flat, accounting for 72\% of the total variability. The second component (26\%) was suppressed below $\sim$ 0.7 keV, then switched to negative at $\sim$ 1.3 keV, and finally became flat above $\sim$ 2 keV. For the third component (0.4\%), it cross the zero-line to the negative side at $\sim$ 0.8 keV and then returns back to the positive side above $\sim$ 7 keV. The fourth component accounts for only 0.3\% of the total variability and is featureless, fluctuating around zero-line in the whole energy range. Comparison with the PCs from the real data indicates that the first simulated PC is consistent with the one from real data, while the second simulated PC significant deviates away from the real one. Specifically, the slope of the PC from simulation shows clear inconsistence with the one from the real data in the whole energy band. 

In panel D, it is the simulated PCs when variation of the free parameters in the absorption model {\sc zxipcf} and {\sc zedge} are considered. We found that the absorptions have very limited influence in the overall shape of the first and second component. The main effect is that the high energy end of the first PC and the $\sim$ 0.8 keV part of the second component get a little suppressed. While the absorptions have clear influence in the third component, suppressing the variability below $\sim$ 0.8 keV and enlarging the variability in $\sim$ 5-8 keV.

\subsubsection{Reflection model}

\begin{figure*}
\center
\includegraphics[width=0.35\textwidth]{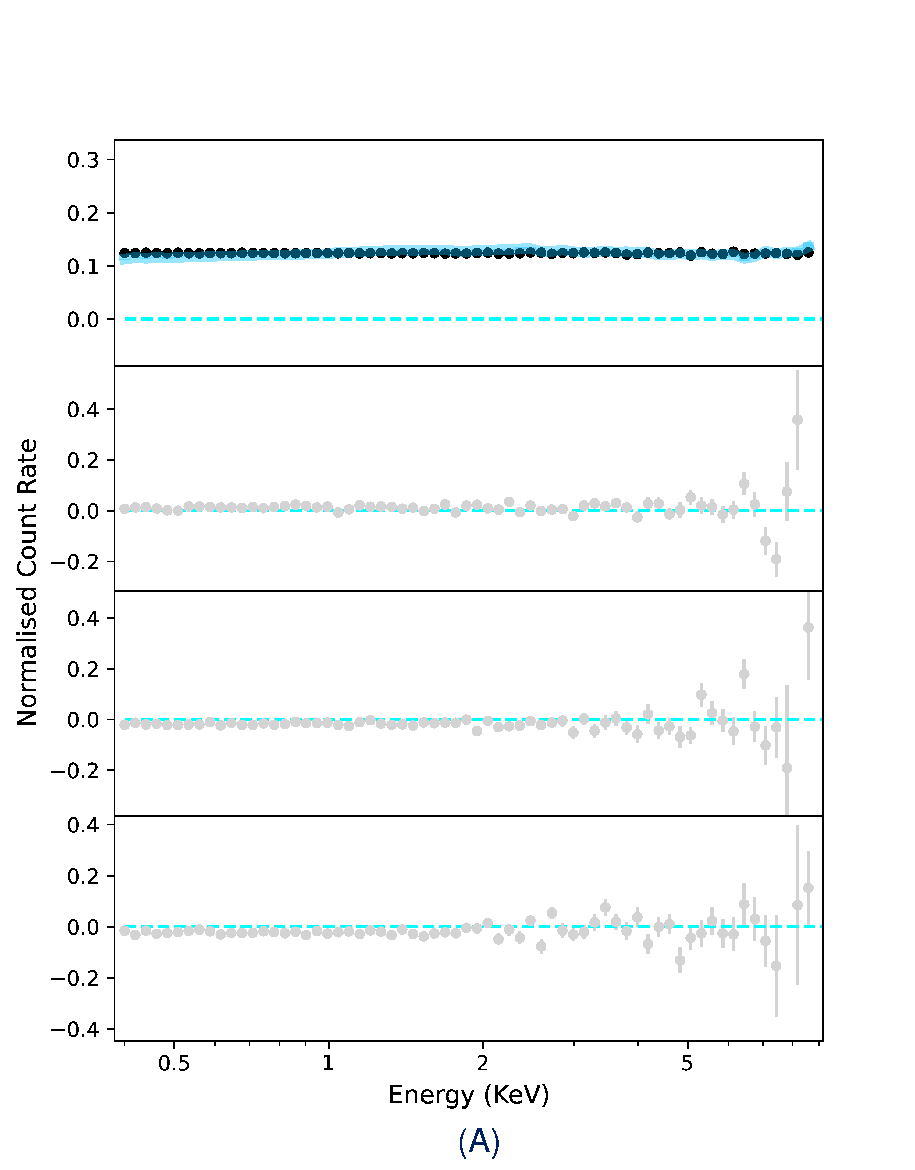}
\includegraphics[width=0.35\textwidth]{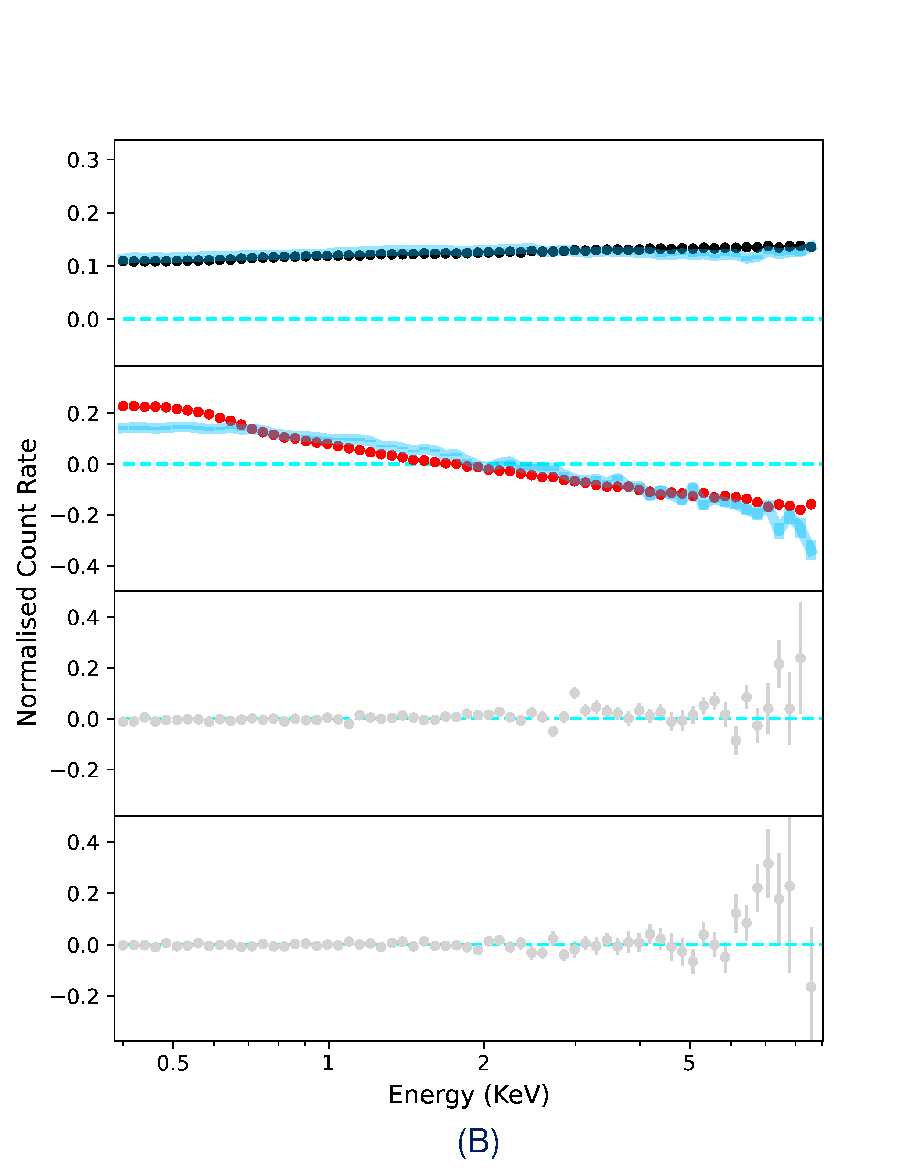}
\includegraphics[width=0.35\textwidth]{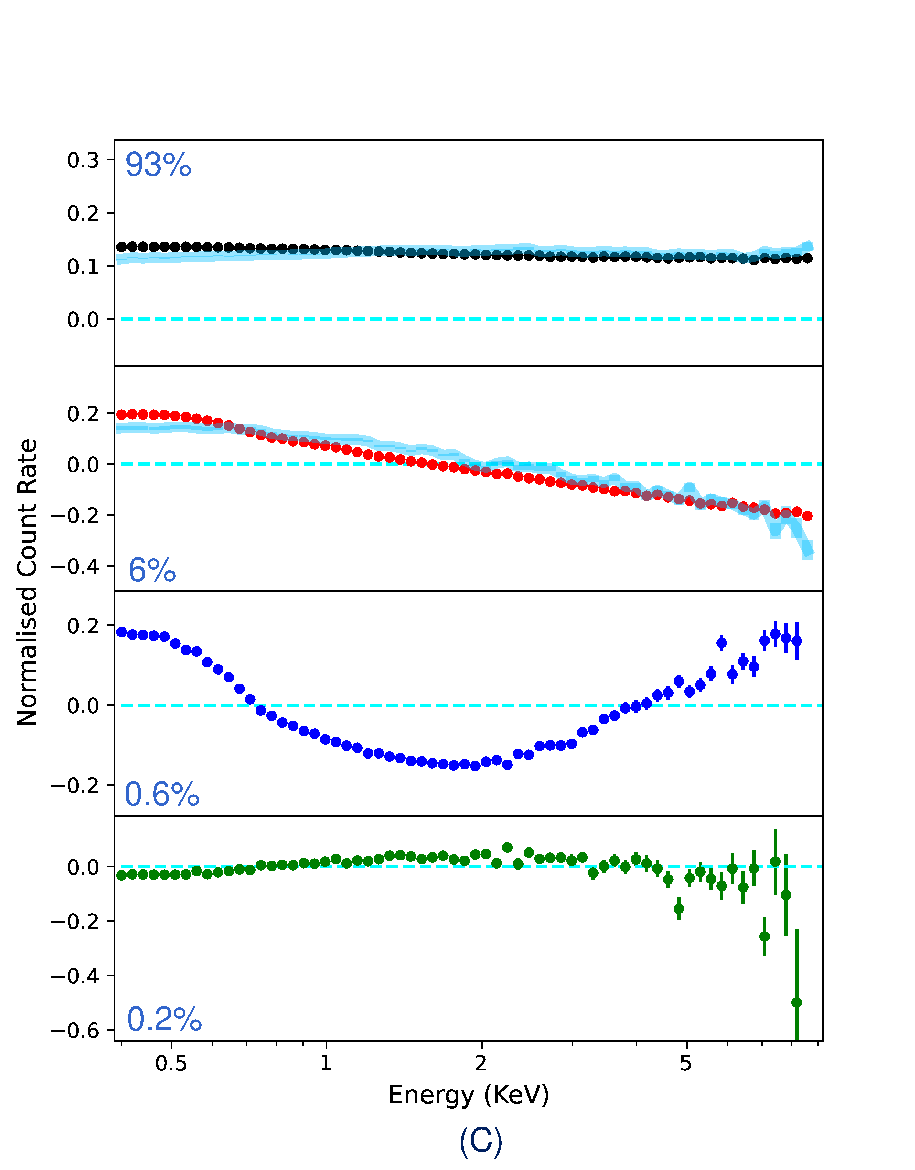}
\includegraphics[width=0.35\textwidth]{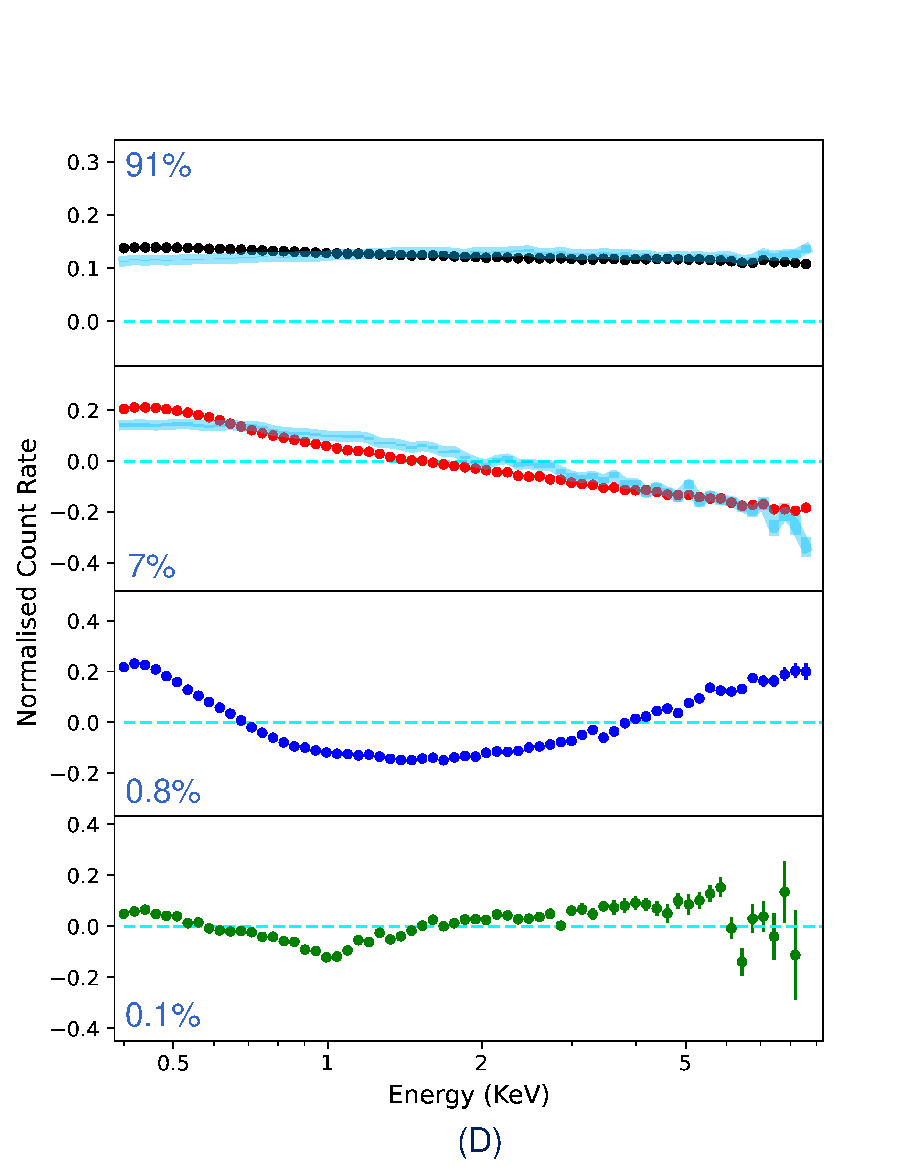}

\caption{Simulated PCs from the variations of different parameters in the relativistically smeared reflection model. When we first vary the parameter Nor$_{relxillCp}$, there is a significant PC above zero (panel A). The second pivoting component comes out when we further set the power-law index $\Gamma$ free to change (panel B). In panel C, we show the four components in the case that log($\xi$), log(N), and $R_{refl}$ in the continuum {\sc relxillCp} are also set free to vary. We finally change the free parameters in the absorption model {\sc zxipcf} and {\sc zedge} and show the corresponding PCs in panel D. In the figure, we used cyan curves to show the first two PCs from the real data for comparison.}
\label{simu_ref}
\end{figure*}

In Figure \ref{simu_ref}, it is the simulation results from the reflection model. As shown in panel A, the first component arised from the variation of the Nor$_{relxillCp}$ always keep flat above the zero-line, matching very well with the one from the real data. The second component appears (panel B) when the variation of the parameter $\Gamma$ is also included. It is flat below $\sim$ 0.6 keV and then behaves like a "straight line" crossing the zero-line at around 1.5 keV.  

In panel C we show the simulated PCs when the change of the continuum parameter log($\xi$), log(N) and $R_{refl}$ are further included. In this case the first principle component is very flat, and it accounts for 93\% of the total variability. The second component (6\%) is flat below $\sim$ 0.6 keV and cross the zero-line around 2 keV. For the third component (0.6\%), it deviates from zero-line in 0.4-0.5 keV, and then it comes to the other side of the zero-line, finally it cross the zero-line again at $\sim$ 5 keV. The 4th component (0.2\%) shows no clear feature and keeps to be consistent with the zero-line below $\sim$ 8 keV. As shown in the figure, the simulated first and the second PC are generally consistent with the ones from the data. In panel D, we show the PCs when absorption effects are included. Apparently, the absorption has little influence in the first and the second component, which contributes the overwhelming majority ($\ge$ 98\%) of the total variability in simulations.


\section{Discussion} 
In this work, we study the variability patterns in Ark 564 with the PCA method using \textit{XMM-Newton} observations. We found that the source show stable variability components (PCs) in different epochs spanning $>$ 10 years. More importantly, although its spectra could be described by both the warm corona model and the reflection model, its PCs favor the reflection scenario as it could well reproduce the observed PCs from real data. As a comparison, the warm corona model fails to reproduce the second principle component derived from the data in Ark 564.

The soft excess origin is still an open question and the origin in different sources may be different \citep{ding22}. Previous spectroscopies have showed different scenarios for the soft excess in Ark 564. \citet{crummy06} studied \textit{XMM-Newton} observations of many type 1 AGNs including the Ark 564. They found that a relativistically blurred photoionized disc reflection model could successfully reproduce the continuum shape, including the soft excess. \citet{brink07} showed that the soft excess flux in Ark 564 is significantly variable on a timescale as short as $\sim$ 0.5-1 ks, and the soft excess could corresponds to the disk reflection emission. Conversely, \citet{dewangan07} found that photons in 4-10 keV energy band lag behind the ones in 0.2-0.5 keV by $\sim$1.8 ks, thus excluding the possibility that the soft excess in Ark 564 is the reprocessed emission of the primary X-ray spectrum. They proposed that the soft excess originates from an optically thick corona in addition to the hot one for the power-law emission. This scenario is further supported by the work of \citet{sarma15} and \citet{ezhikode16}, in which they reported that the X-ray spectra of Ark 564 could be well fitted by a double Comptonization model: one for the soft excess and the other for the hard X-ray power law component.

\citet{parker15} studied 26 AGNs with the PCA method including Ark 564 and built a library of different PCs in order to quickly distinguish variability behaviors in AGNs. For the soft excess and the hard excess present in the third component from the PCA, \citet{parker15} proposed that it could not due to a Comptonization or bremsstrahlung process in the warm corona since they cannot extend to high energies to generate a simultaneous hard excess. In their simulations, due to the lack of information on the physical parameters, a simple flux ratio has been assumed between the different components, with the varying factor of each flux being assigned to certain values. 

To make comparisons with their work, in this study, we combine the results from spectral analysis. Hence, we were able to obtain the variation ranges of the related physical parameters accurately. This is important since the shape and the relative strength of the features in the simulated PCs are, to a great extent, influenced by the variation ranges of the physical parameters. Therefore, by incorporating the parameter ranges from the spectroscopy, the PCA can more realistically reflects the consistency between the variability pattern of the real data and the models.


Our simulations indicate that the warm corona model fails to generate the observed pivoting component (the second PC). Compared with the one from the real data, the second principle component that the warm corona model produces has a different shape and origin. We found that this simulated pivoting PC is due to the variations in the two normalization parameters. This is different from the origin of the one from the real data (Figure \ref{pca}), which is believed to be caused by a power law varying in its photon index $\Gamma$ \citep[see, e.g.,][]{parker14a,parker15}. 

Apart from the origin of the reflection, recent studies have show that a combination of the relativistic reflection and the warm corona likely be a natural scenario responsible for the soft excess \citep[see, e.g.,][]{porquet18,porquet21,xu21}. Theoretically, \citet{xiang22} proposed a hybrid scenario which self-consistently combines the effects from the reflection and the warm corona. According to their model, the accretion energy released in the inner disk is distributed between a warm corona, a hot corona, and the disk. As the fraction of the energy dissipated in the warm corona and the hot corona varies, the soft excess shows variety of shapes and sizes. For the case of Ark 564, \citet{lewin22} simultaneously fitted its time lags and the flux spectra with \textit{XMM-Newton} and \textit{NuSTAR} observations. Their analysis showed that a blackbody component in addition to the disk reflection component is required to describe the flux spectra. The blackbody component represents the Comptonization off a warm corona, so both the warm corona and the relativistic reflection likely contribute to the soft excess, similar to the scenario proposed in the "hybrid" model.

It could be also possible that the soft excess is not connected to the warm corona. The warm corona is proposed to be a Comptonizing regions exist at the surfaces of the accretion disk in AGNs. Nowadays there are still concerns about its physical plausibility. The theoretical work by \citet{rozanska15} and \citet{gronk20} showed that strong magnetic pressure support is a prerequisite in order to generate a $\tau$ $\sim$ 10 warm corona in hydrostatic equilibrium. \citet{ballantyne20} demonstrated that only when the gas densities and the temperatures in a limited range could the warm corona generate a smooth soft excess, indicating that the warm corona must be located close to the hot corona region in order to reach sufficient high ionization. In addition, \citet{gronk20} found that the formation of the warm corona is prevented by thermal instability in the case of a low accretion rate. This statement is supported by \citet{ballantyne20b}, which indicates that system with a low accretion rate is unable to provide enough energy to sustain a warm corona. Although the accretion rate in Ark 564 is high, making the aforementioned scenario unlikely, there is still no clear evidence to prove the existence of the warm corona in this system in the current stage.   

In our fitting results there are several absorption edges, which likely reflect elements and their charge states in the absorber. In the case of relativistically blurred reflection, the energy range for the four edges is 0.3-0.42 keV, 0.509-0.54 keV, 1.0-1.19 keV, and 1.17-1.43 keV, respectively. Absorption edges at similar energies in this source are also reported in the work of \citet{sarma15}. By comparing these edges to the identification results in \citet{turner04}, we found that the edges below 0.5 keV could be due to the C V, C VI, S XIII, Ar XII, and Si XII, and the edges around 0.52 keV, 1.1 keV, and 1.3 keV have mainly arisen from the element O I-II, Ne X plus Fe XXIII, and Mg XI. 

These narrow absorption features suggests that the soft excess in Ark 564 is unlikely to have been caused by the complex absorptions. \citet{gierlinski04} proposed that the soft excess would be due to strong, relativistically smeared, partially ionized absorption, which may come from a differentially rotating, outflowing disc wind. In this scenario, there should be no narrow absorption lines present in the observed spectra since they are smeared and reshaped as a "quasi-continuum" by a very large dispersion velocity. Obviously, the detection of the narrow absorption edges in this work disagrees with the absorption origin of the soft excess in this source. This finding is also consistent with the conclusion in the work of \citet{dewangan07}, where they found that a unphysically large smearing velocity ($\sim$ 0.8c) is required in Ark 564 in order to generate the observed smooth soft excess emission by the absorptions .

In this work, we show that the PCA method is capable of distinguishing different physical interpretations of the soft excess in AGNs, and that it can be applied to other NLSy1s and BLSy1s for their variability patterns and the soft excess origin. Furthermore, the PCA together with the spectroscopy would be a powerful tool for the exploration of the complicated "hybrid" warm corona,  together with a blurred reflection scenario in the future.

\section*{Acknowledgement} 
We thank the anonymous referee for his/her careful reading of the manuscript and useful comments and suggestions. This research has made use of data obtained from the High Energy Astrophysics Science Archive Research Center (HEASARC), provided by NASA's Goddard Space Flight Center. This research made use of NASA's Astrophysics Data System. We thank M. L. Parker for the PCA code support. Lyu is supported by Hunan Education Department Foundation (grant No. 21A0096). X. J. Yang is supported by the National Natural Science Foundation of China (NSFC 12122302 and 12333005). Z.Y.F. is grateful for the support from the Postgraduate Scientific Research Innovation Project of Hunan Province (grant No. CX20220662) and the Postgraduate Scientific Research Innovation Project of Xiangtan University (grant No. XDCX2022Y072).

\bibliographystyle{aa}
\bibliography{biblio}
%



\end{document}